\documentclass[twocolumn]{aastex631}

\begin{document}

\title{Blowing star formation away in AGN Hosts (BAH) - I.  Observation of Warm Molecular Outflows with JWST MIRI}

\author[0000-0003-3667-9716]{J. H. Costa-Souza}
\author[0000-0003-0483-3723]{Rogemar A. Riffel}
\author[0009-0005-0583-5773]{Gabriel L. Souza-Oliveira}
\affiliation{Departamento de F\'isica, CCNE, Universidade Federal de Santa Maria, Av. Roraima 1000, 97105-900,  Santa Maria, RS, Brazil}

\author[0000-0001-6100-6869]{Nadia L. Zakamska}
\affiliation{Department of Physics \& Astronomy, Johns Hopkins University, Bloomberg Center, 3400 N. Charles St, Baltimore, MD 21218, USA}

\author[0000-0002-6570-9446]{Marina Bianchin}
\affiliation{Department of Physics and Astronomy, 4129 Frederick Reines Hall, University of California, Irvine, CA 92697, USA}

\author[0000-0003-1772-0023]{Thaisa Storchi-Bergmann}
\author[0000-0002-1321-1320]{Rog\'erio Riffel}
\affiliation{Departamento de Astronomia, IF, Universidade Federal do Rio Grande do Sul, CP 15051, 91501-970, Porto Alegre, RS, Brazil}



\begin{abstract}
 We use the James Webb Space Telescope (JWST) Mid-Infrared Instrument (MIRI) medium-resolution spectrometer (MRS) observations of the radio loud AGN host UGC\:8782 to map the warm molecular and ionized gas kinematics.  The data reveal spatially resolved outflows in the inner 2 kpc, seen in low ionization (traced by the [Ar\,{\sc ii}]\,6.99\,$\mu$m emission) and in warm molecular gas (traced by the H$_2$ rotational transitions). We find a maximum mass-outflow rate of $4.90\pm2.04\:$M$_{\odot}$\:yr$^{-1}$ at $\sim$900\,pc from the nucleus for the warm outflow (198\,K$\leq\:$T$\leq\:$1000\,K) and estimate and outflow rate of up to $1.22\pm 0.51$\:M$_{\odot}$\:yr$^{-1}$ for the hotter gas phase (T$\: > \:$1000\,K). These outflows can clear the entire nuclear reservoir of warm molecular gas in about 1 Myr. The derived kinetic power of the molecular outflows lead to coupling efficiencies of 2-5 percent of the AGN luminosity, way above the minimum expected to the AGN feedback be effective quenching the star formation. 
\end{abstract}


\section{Introduction} 

Among the myriad challenges in constructing robust models of galaxy evolution, one glaring gap persists: the scarcity of galaxies at both ends of the mass spectrum during the later stages of cosmic history. This puzzling phenomenon presents a significant enigma in extragalactic astrophysics: Why are there so few galaxies of both low and high masses? The absence of galaxies at both mass ends indicates that our understanding of galaxy evolution remains incomplete, with numerous factors yet to be fully explored or discovered. 
{Among these factors, supernova \citep{2015Natur.523..169E,2017MNRAS.469.4831C} and ionization feedback \citep{2010arXiv1009.4505B,2013ApJ...765...22B} emerges as crucial components, essential for suppressing the formation of stellar mass in galaxies at the low-mass end. At the high-mass end, supernova feedback is insufficient \citep{2011IAUS..277..273S}, but the energy demands for removal or reheating the gas available for star formation can potentially be met by Active Galactic Nuclei (AGN)}. Indeed, numerous models have been developed to elucidate AGN feedback, demonstrating that even with low coupling efficiencies, 
AGN can exert a substantial impact on their host galaxies \citep[e.g.][]{2017MNRAS.464.2840A,2023MNRAS.526..217A}. AGN-driven winds have the ability to reheat, expel or redistribute the gas within the galaxy. Because the AGN-driven winds are a multi-phase phenomenon, with a wide range of it is imperative to thoroughly examine each.

The molecular gas is the main driver of star formation (SF) \citep{2012ARA&A..50..531K} and AGN activity \citep{2019NatAs...3...48S}. Conventionally, it is common to separate molecular gas into three phases with well-defined tracers:
 (i) The cold gas, in which the temperatures reaches no more then 100\,K. This phase is the densest gas phase and is commonly traced by (sub)millimeter transitions of molecules such as CO and HCN; (ii) The warm molecular gas phase, 100\,K\,$\lesssim T \lesssim$ 1000\,K, summing up to $\sim$ 1-30 percent of total molecular gas mass \citep{2007ApJ...669..959R}. This specific gas phase can be traced by the rotational H$_2$ transitions seen in the mid-infrared (MIR); (iii) The hot molecular phase, $T\,\gtrsim1000\,$K, mainly traced by the roto-vibrational H$_{2}$ transitions in the near-infrared. This gas phase represents only $10^{-7}-10^{-5}$ of the mass of the cold molecular gas phase \citep{2005AJ....129.2197D,2014A&A...572A..40E,2016A&A...594A..81P}. Spatially resolved observations of the structure and kinematics of both cold and hot molecular gas emissions in the central regions of galaxies have been meticulously obtained through ALMA and near-infrared integral field unit observations \citep{2022A&A...668A..45L,Ramos-Almeida22,Speranza22,Bianchin22,Rogemar_21_Cyg,2023MNRAS.521.1832R}. 
 
 With the advent of JWST, it is now possible to conduct detailed studies of the warm molecular phase, in particular with the Mid-Infrared Instrument (MIRI) medium-resolution spectrometer (MRS).   These studies reveal a wide range of interactions between different gas phases, as found for the Stephan's Quintet, one of the first low-redshift system to undergo a detailed analysis \citep{2023ApJ...951..104A}; molecular gas tracing the deceleration of the radio jets by interaction with the disk in NGC\,7319 \citep{2022A&A...665L..11P}; detection of highly ionized outflows in a Compton-thick AGN \citep{2023A&A...672A.108A}; the presence of warm molecular outflows in NGC\:3256, a nearby starburst galaxies \citep{2024arXiv240314751B}, among others \citep[e.g.][]{2022A&A...666L...5G,2023ApJ...948..124H,2023ApJ...942L..37A,2023arXiv231204914G,2024arXiv240106880F,2024ApJ...960..126V,2022ApJ...940L...5U}.
 
Spitzer telescope observations of nearby galaxies have revealed a significant excess of H$_2$ emission observed from the rotational transitions in AGN hosts compared to star-forming galaxies \citep{2019MNRAS.487.1823L}. 
 Furthermore, the MIR H$_2$ emission lines are especially conspicuous in shocked gas \citep{2014MNRAS.439.2701H} and are predicted by numerical simulations to be valuable tools for investigating the warm molecular gas phase of galactic outflows \citep{richings18a,richings18b}. Moreover, the surplus of H$_2$ emission observed in AGN hosts correlates with optical shock tracers, such as [O\,{\sc i}]$\lambda$6300, offering observational evidence that H$_2$ MIR lines can be employed to observe gas shocked by AGN winds \citep{2020MNRAS.491.1518R}. Besides the H$_2$ transitions, the MIR spectral range includes also shock tracers in partially ionized gas zones, such as [Fe\,{\sc ii}] emission lines \citep{2004A&A...425..457R,2013MNRAS.430.2002R,2018ApJ...866..139K}, and emission lines from the fully ionized gas.

 Here, we use JWST MIRI observations to map the gas kinematics in the 3.5\,$\times\,$3.5\,kpc$^{2}$ central region of the radio-loud AGN, UGC\,8782 (3C 293). This galaxy was selected from the sample of \citet{2020MNRAS.491.1518R} as one of the objects with the largest H$_2$ emission excess (traced by the H$_2$S(3)\,9.665\,$\mu$m/PAH\,11.3\,\,$\mu$m emission-line ratio) as well as the highest [O\,{\sc i}]$\lambda$6300 velocity dispersion and [O\,{\sc i}]$\lambda$6300/H$\alpha$ flux line ratio in their sample. This makes this galaxy an excellent candidate to host warm molecular gas outflows.  UGC\,8782 is located at a distance of 203.4$\pm$\,14.21 Mpc \citep{2017ApJS..233...25A}, hosting a radio loud AGN \citep[e.g., ][]{liu02}, and a double-double radio source with the outer lobes extending up to 200\,kpc  the northwest-southeast direction and inner lobes on a $\sim$4 kpc scale aligned along the east-west direction \citep[e.g., ][]{machalski16}. Based on optical diagnostics, its nuclear activity is classified as a Low-Ionization Nuclear Emission Region \citep[LINER, ][]{veron06} and optical images reveal dust filaments on scales from 100s of parsecs to a few kpc, associated with a previous merger event \citep[e.g.,][]{martel99}. UGC\,8782 displays multi-gas phase outflows, observed in neutral hydrogen  \citep{2003ApJ...593L..69M,2013MNRAS.435L..58M}, and in ionized gas \citep{2005MNRAS.362..931E,2016MNRAS.455.2453M,2023MNRAS.521.3260R}.  These outflows are interpreted as being mechanically driven by the radio jet and the kinetic power of the ionized gas outflows is in the range of 1--3 per cent of the AGN bolometric luminosity \citep{2023MNRAS.521.3260R}. The cold molecular component of the outflows remains undetected thus far \citep{labiano14} and here we report, for the first time, the observations of warm molecular outflows by studying the mid-infrared H$_2$ emission, using spatially resolved observations with JWST.

This paper is organized as follows: In Sec.~\ref{sec:data} we describe the data and measurements, while Sec.~\ref{sec:results} presents the results and discuss them. Sec.~\ref{sec:conclusion} summarizes our conclusions.

\section{Data and Measurements}\label{sec:data}

\begin{figure*}
    \centering
    \hspace{-1cm}\includegraphics[width=0.88\textwidth]{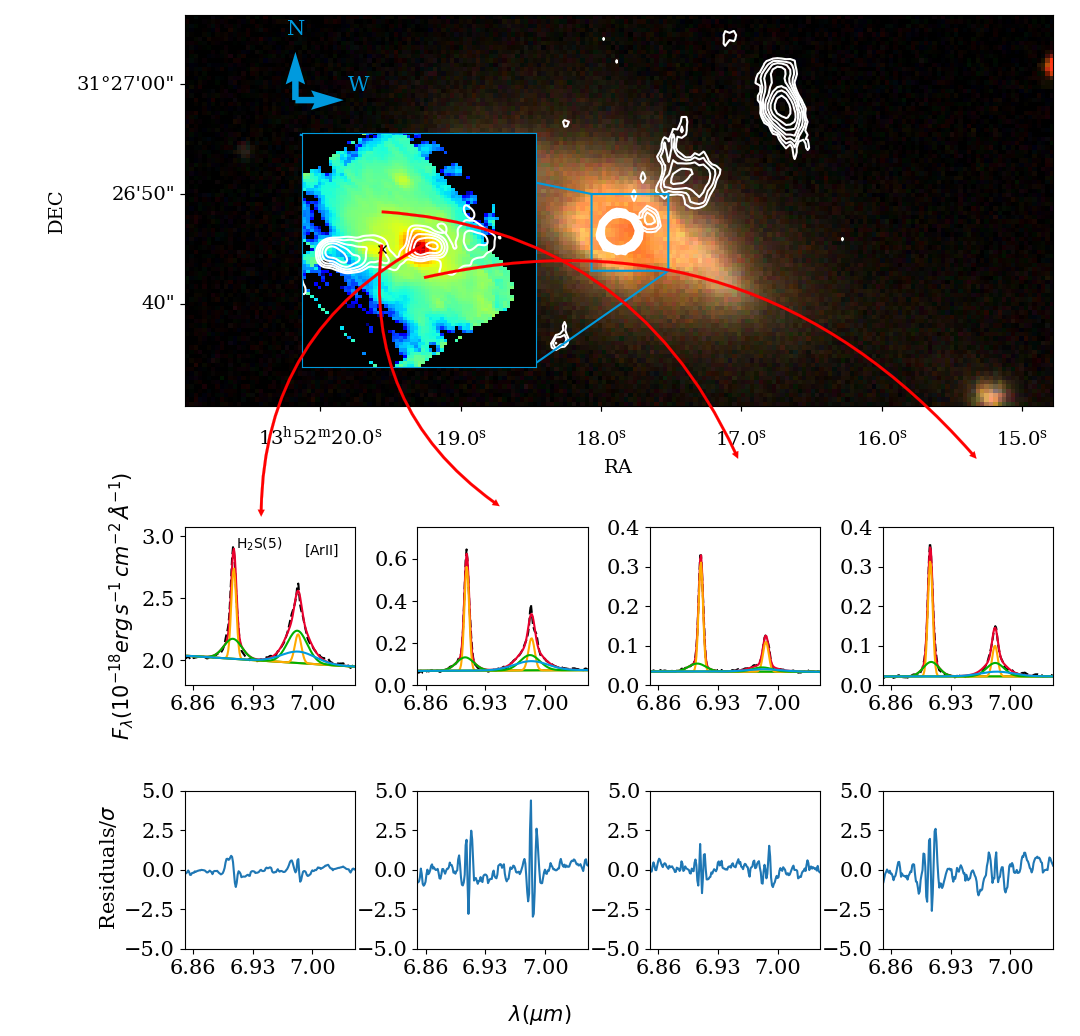}
        \includegraphics[width=0.88\textwidth]{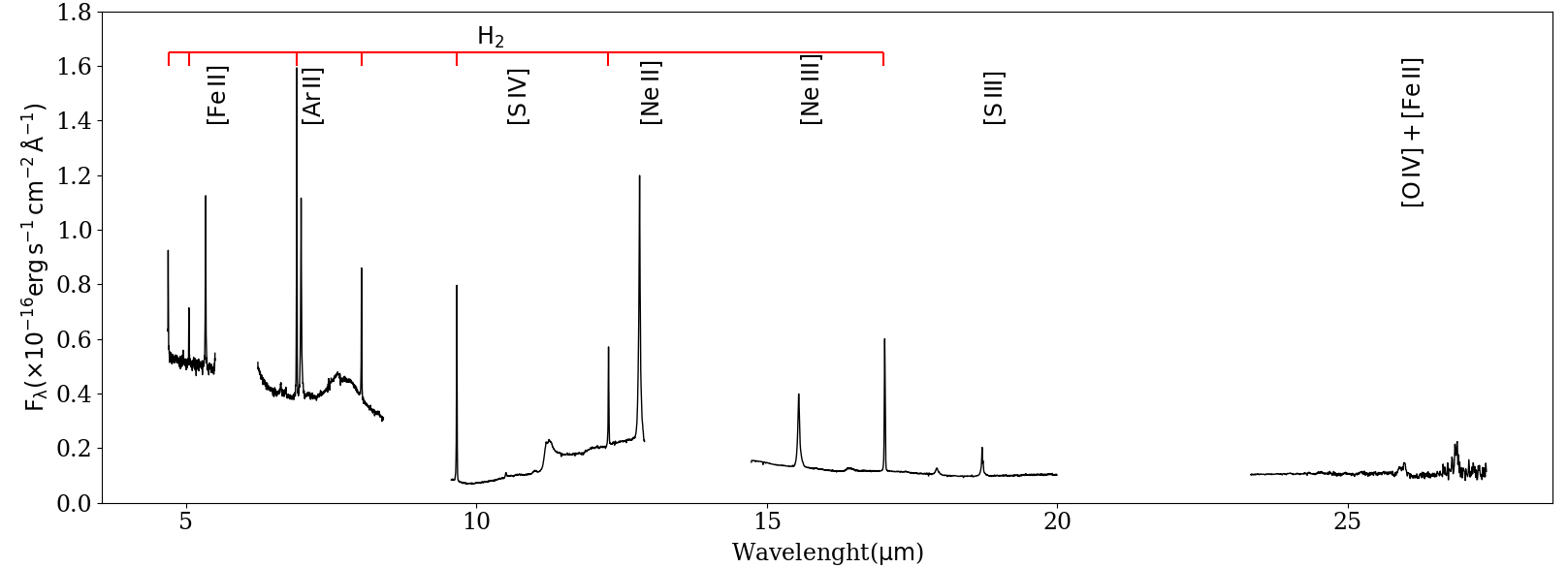}
        \caption{Sloan Digital Sky Survey RGB composition of UGC\,8782. The white contours trace the radio-jet at 1.4GHz, obtained from VLA archive (\cite{1995ApJ...450..559B}). The zoomed in panel displays the JWST MIRI 7.0\,$\mu$m continuum image. The white contours are from a L-Band (1.65\,GHz) image from the e-Merlin archive. The position of the MIR emission peak is indicated by a $+$ sign, while the $\times$ shows the position of the optical nucleus.  The plots show examples of fits of the emission lines for four positions, along with the residuals of the fits divided by the standard deviation of the nearby continuum.  The bottom panel shows an integrated spectrum within a circular aperture of 1\,arcsec around the nucleus. }
    \label{fig:large}
\end{figure*}

The observations were performed with the JWST using the \citep[Mid-Infrared Instrument (MIRI)][medium-resolution spectrometer (MRS)]{2015PASP..127..646W,2023A&A...675A.111A} in the sub-bands Short and Long.  We adopted a four point dither pattern, with 30 groups and a single integration for each dither position. We used the slow readout mode, which yields a total exposure time on source of 1\,h\,35m, and we obtain an additional 24\,m of background exposure performed at the first dither position. 

The science and background observations were processed using the JWST Science Calibration Pipeline \citep{bushouse_2024}, version 1.13.4, and reference file \texttt{jwst\_1188.pmap}. After completing the initial data reduction steps, we process the cosmic ray files  through a custom outlier detection script to remove residual outliers. To eliminate the resampling noise, a necessary step to fit the emission lines \citep{law23}, we reassign the spectrum of each spaxel to the integrated light within an aperture of 0.5 arcsec centered on the spaxel. The resulting spectrum is then renormalized by multiplying it by the ratio of the observed continuum to the integrated spectrum's continuum at each wavelength. This approach does not significantly affect the amplitude and kinematics of the emission lines. A similar procedure is adopted by \citet{perna23} for processing NIRSPEC IFU data. {The Background subtraction was performed using the master background subtraction method. In this process, a 1-D master background spectrum is derived from the background exposures and then projected into the 2-D space of the source data, to perform the subtraction.}
A detailed description of the observations and data reduction procedure  will be provided in a forthcoming paper (Souza-Oliveira et al., in prep.).

The resulting data covers the spectral region 4.7-27$\mu $m, excluding the Medium sub-band windows (5.49-6.24 $\mu$m, 8.39-9.57 $\mu$m, 12.88-14.74 $\mu$m and 20.00-23.36 $\mu$m). The mid-infrared spectrum of UGC\,8782 is abundant in emission lines from the molecular gas (H$_{2}$ rotational transitions), as well as emission lines from ionized species such as [Fe\,{\sc ii}]\,5.34$\mu$m, [Ar\,{\sc ii}]\,6.99$\mu$m, [Ne\,{\sc ii}]\,12.81$\mu$m, [Ne\,{\sc iii}]\,15.56$\mu$m, [S\,{\sc iii}]\,18.71$\mu$m, and a few lines are quite fainter ([Si\,{\sc iv}]\,10.51$\mu$m, [Fe\,{\sc ii}]\,17.94$\mu$m), as observed in the bottom panel of Fig.~\ref{fig:large}. Here, we focus our analysis in the H$_{2}$  rotational transitions, tracers of the warm molecular gas, and use the [Ar\,{\sc ii}]\,6.99$\mu$m as a tracer of the ionized gas emission. The angular resolution is about $\sim$0.36 arcsec as obtained by measuring the FWHM of the flux distributions of the stars HD\,192163 and HD\,2811 at 7.0\,$\mu$m, after processing the data following the same procedure used to obtain the datacube for UGC\,8782.

We use the IFSCUBE Python package \citep{Ruschel-Dutra21} to fit the emission lines by Gaussian functions for each profile. {We create the observed continuum by applying a double lowess smoothing for each spectral interval in the observed spectrum, using 30\% and 10\% of the data points in this order, followed by linear interpolation. Then, the emission lines are fitted using the continuum subtracted spectra.} The H$_2$ lines are well reproduced by two Gaussian each, while the [Ar\,{\sc ii}]\,6.99$\mu$m, required an additional component to reproduce the line profile. The residuals from the fits for all lines are comparable to the noise in the continuum. They exceed 5$\sigma$ of the continuum noise level in only 5\% of the spaxels with measurements for H$_2$ S(5) and 3\% for the [Ar\,{\sc ii}].
Examples of the fits are shown in Fig.~\ref{fig:large} for four positions.  The IFSCUBE code outputs a data cube including the best-fit parameters which are used to map the emission line flux distributions and kinematics.

\begin{figure*}
    \centering
    \includegraphics[width=.7\textwidth]{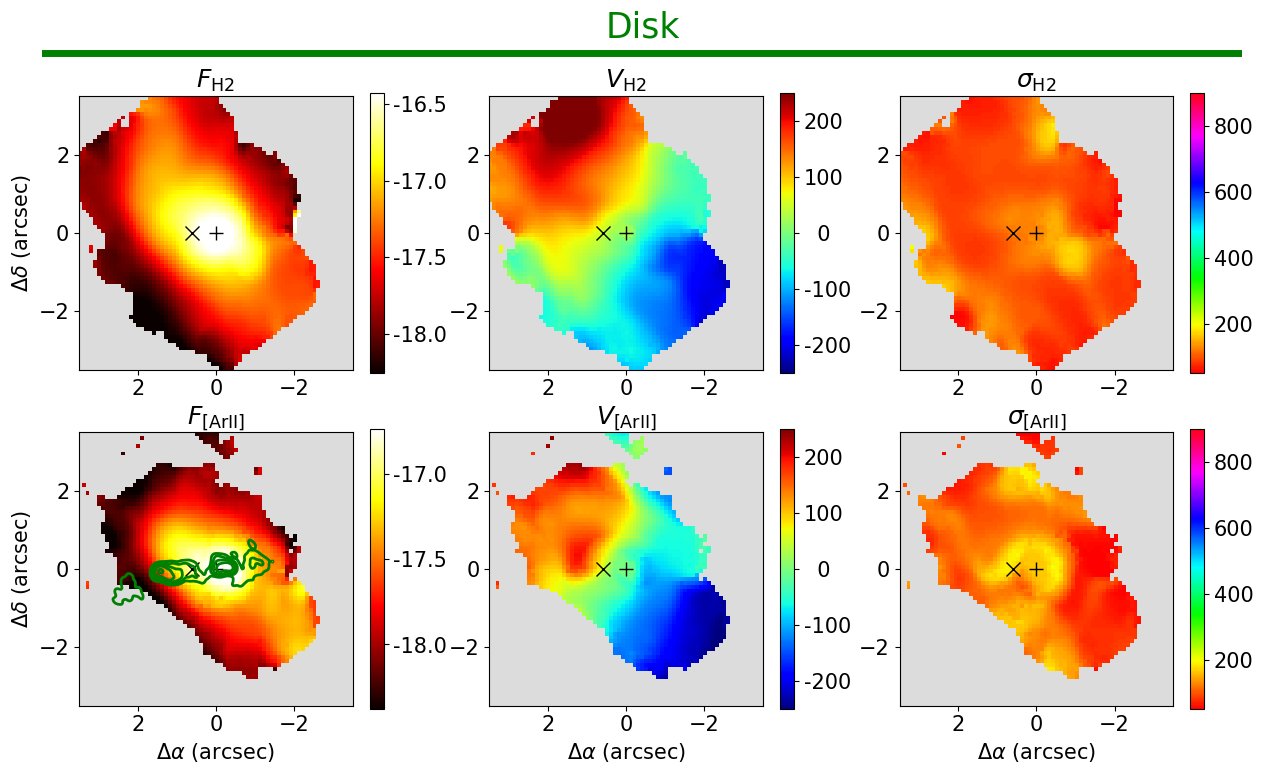}
    \includegraphics[width=.7\textwidth]{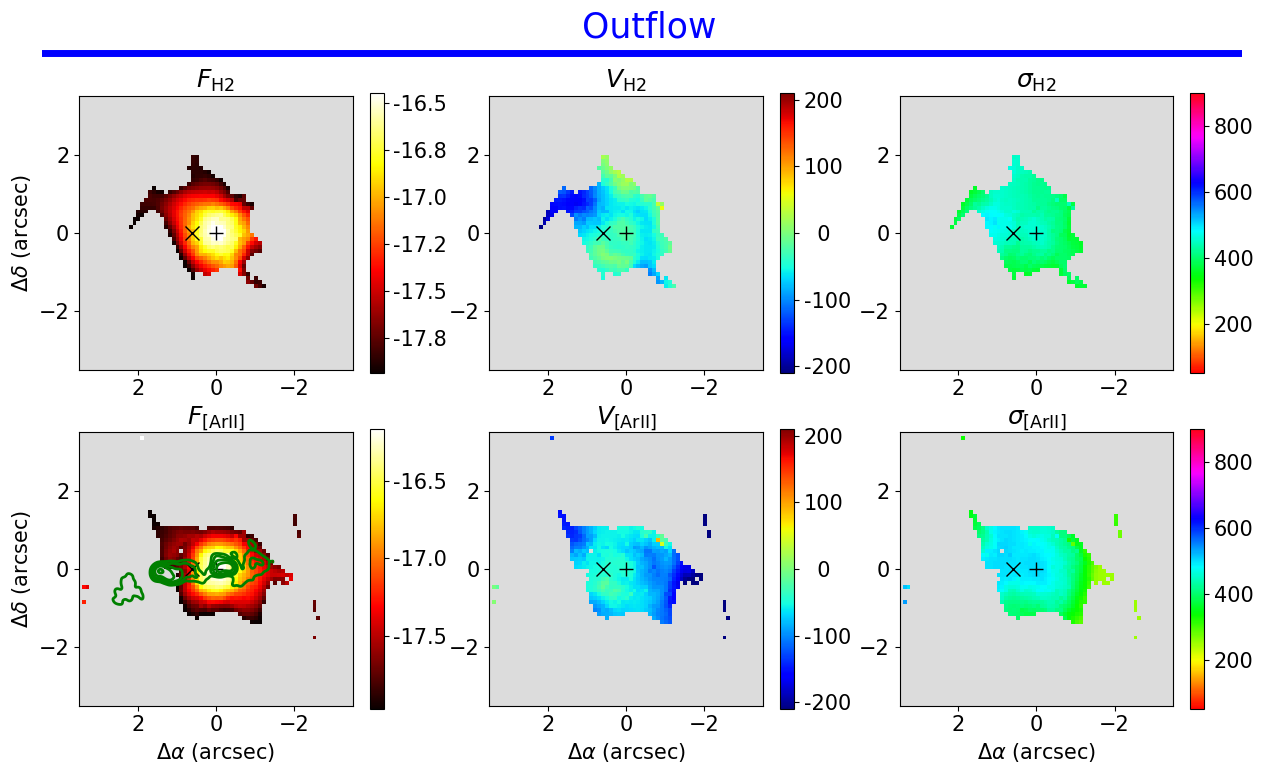}
    \includegraphics[width=.7\textwidth]{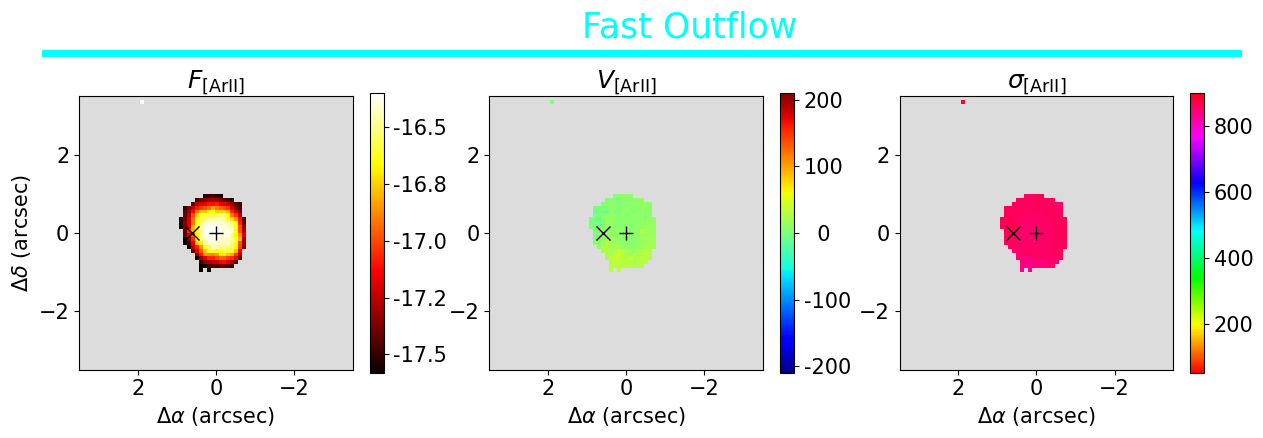}
    \caption{The top panels show the flux distribution in log$_{10}(\rm erg\,s^{-1}cm^{-2}$) (left), velocity field (center) and velocity dispersion map (right) for the narrow (disk) component of the H$_{2}$ S(1) (top)  and [Ar\,{\sc ii}]\,6.99$\mu$m (bottom) emission lines. The $+$ and $\times$ symbols mark the position of the MIR and optical continuum peaks, respectively. The gray regions correspond to locations where the emission lines are not detected with at least 3 sigma above the continuum noise. The green contours are from the L-Band (1.65\,GHz) image from the e-Merlin archive. The  middle panels show the results for the outflow component with the same nomenclature used for the disk component,  while the bottom panels show the results for the fast outflow seen in [Ar\,{\sc ii}]\,6.99$\mu$m. } 
    \label{fig:3}
\end{figure*}

\section{Results and discussions} \label{sec:results}

We find that the peak of the continuum emission in the MIR is shifted by $\sim$0.6 arcsec to the west in relation to the peak of optical emission, while the emission peaks of optical lines and the IR are co-spatial \citep{2023MNRAS.521.3260R}. 
The MIR emission peak is consistent with the position of the radio core.
This is consistent with previous observations that reported a displacement between the optical center and the radio core in UGC\,8782 \citep[e.g.][]{2005MNRAS.362..931E,2016MNRAS.455.2453M} and is likely due to higher extinction in the optical continuum.

In Fig.~\ref{fig:3}, we present the flux distributions and kinematic maps for the H$_{2}$S(5)\,6.91$\mu $m and [Ar\,{\sc ii}]\,6.99$\mu$m  emission lines.\:The top panels show the results for the narrow component, attributed to the disk emission, while the  middle panels show the resulting maps for the broad component, due the outflow. In addition, a very broad component is observed in [Ar\,{\sc ii}]  (a fast outflow), with a median velocity dispersion of $\sigma\sim$860\,km\,s$^{-1}$ and  centroid velocity close to galaxy's systemic velocity. We do not show the maps for this component as it is very compact, extending to up to 0.8 arcsec from the MIR continuum peak,  as seen in the bottom panels of Fig.~\ref{fig:3}. The disk component has extended emission over the whole MIRI-MRS field of view and gas kinematics consistent with a rotation pattern centered on the MIR nucleus.

The emission of the outflow component (bottom panels of  Fig.~\ref{fig:3}) is spatially resolved (with FWHM of its flux distribution of about 2.4 times that of the PSF) and concentrated mostly around the MIR nucleus and radio core. In addition, we simulated a rotating disk convolved with the PSF to assess the impact of beam smearing on the observed line profiles, which was found to be negligible (refer to the Appendix~\ref{beam}). This component is blue-shifted relative to the systemic velocity of the galaxy and presents $\sigma$ values of up to $\sim$600\,km\,s$^{-1}$. The median velocity and $\sigma$ values are $-$70\:km\,s$^{-1}$ and 455\:km\,s$^{-1}$ for the [Ar\,{\sc ii}] and $-$50\:km\,s$^{-1}$ and 430\:km\,s$^{-1}$ for the H$_2$, respectively. 
From the observed H$_2$ emission and kinematics, we derive the warm molecular gas outflow properties in UGC\,8782.

We measure the gas mass and temperature using the transitions H$_2$ (0-0) S(1),H$_2$ (0-0) S(2) , H$_2$ (0-0) S(3), H$_2$ (0-0) S(4) and H$_2$ (0-0) S(5), as our data does not include the higher order transitions, as well as the H$_2$ (0-0) S(0) transition. We correct the observed fluxes for extinction using a visual extinction of $A_{\rm V}=1.7$ mag, the mean value for the outflow component \citep{2023MNRAS.521.3260R} and adopting the extinction law G23 \citep{2023ApJ...950...86G,2009ApJ...705.1320G,2019ApJ...886..108F,2021ApJ...916...33G,2022ApJ...930...15D}.   

\begin{figure}
    \centering
    \includegraphics[width=.50\textwidth]{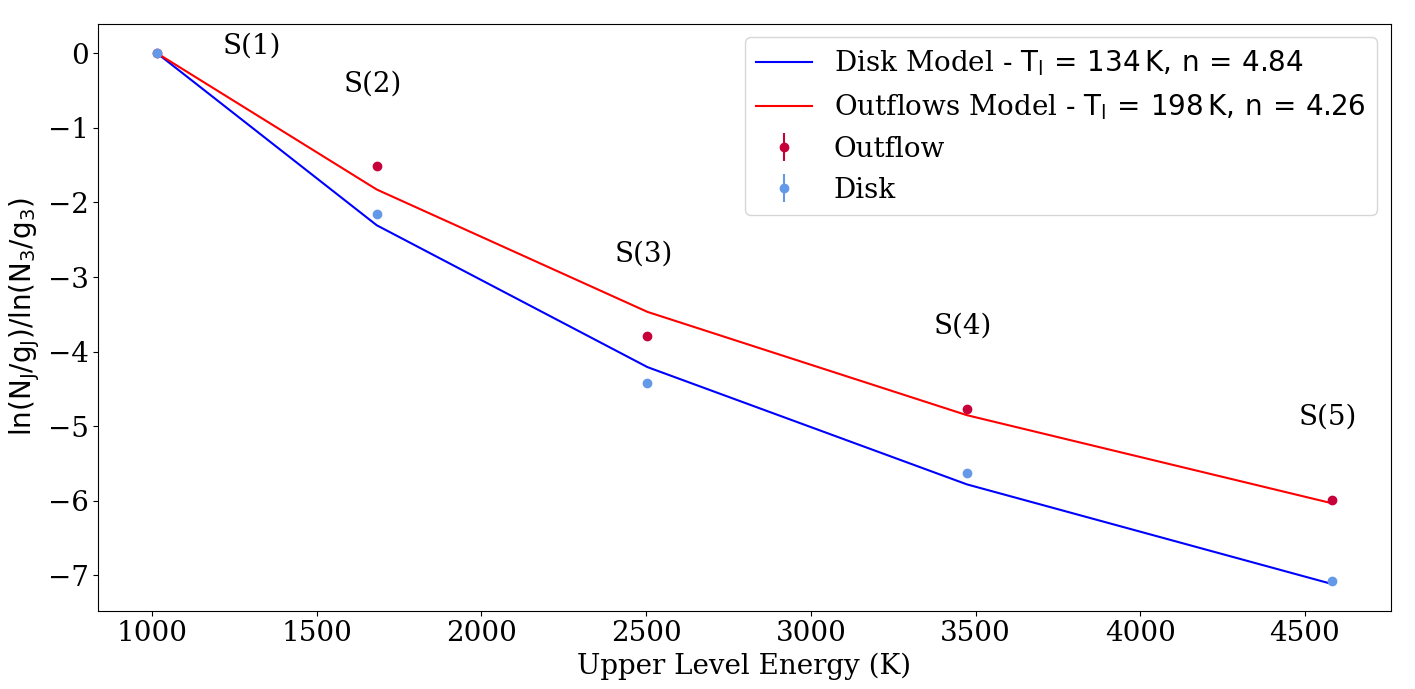}
    \caption{ Excitation diagrams for the outflow and disk components. The {\it y-}axis shows the population number of the {\it J} state divided by the degeneracy of the state and normalized by the {\it J}\,=\,3 state , while the {\it x-}axis shows the upper level energy. The red line, describes the power-law model fitted at the outflow component, while the blue line represents the disk.}
    \label{fig:temp}
\end{figure}

To estimate the mass of the warm molecular gas, we assume that the gas lies in a temperatures distribution, with an optically thin gas. Given that a single temperature is rarely sufficient to reproduce the observed data \citep[e.g.,][]{2007ApJ...669..959R,2006ApJ...648..323H,2018AJ....156..295P}, we employ a power law approach to estimate the mass within a temperature interval $[T, T + dT]$ \citep{2014A&A...566A..49P,togi}. The column density $N_{T}$ can be estimated from the column density of the $J$ state $N_J$, given by
    \begin{equation} \label{eq:NJ}
        N_{J} = \int_{T_{u}}^{T_{l}} f(T)m T^{-n}d T,
    \end{equation}
where $m$ is a normalization constant $m = N_{T}(n-1)/(T_{l}^{1-n}-T_{u}^{1-n})$, $f(T)$ is the Boltzmann factor to the ortho states
\begin{equation}
    f(T)\,=\,\frac{g_{J} e^{{-E_{J}}/{k_{B}T}}}{ \sum_{J'=1}^{\rm ortho} g_{J'} e^{{-E_{J'}}/{k_{B}T}}}.
\end{equation}
$g_{J}= 3(2 J\,+\,1)$ is the degeneracy of the ortho $J$ state and $k_{B}$  is the Boltzmann constant. Following \cite{1991qmnr.book.....L} and \cite{Huber1979} the energy of the state can be obtained by 
\begin{equation}
    E_{J}\,=\,85.35\,{\rm K} \, k_{B} J(J+1)-0.068\,{\rm K}\,k_{B}\,J^{2}(J+1)^2.
\end{equation}
The column density $N_J$ can also be determine directly, by $N_J$ = $4 \pi F_J/\Omega (\Delta E_{u-l}\,A_{u-l})$,  $\Delta E_{u-l}$ is the energy of the transition, $A_{u-l}$ is the Einstein coefficient of the considered transition, and $\Omega$ is the solid angle of observation. Replacing $N_J$ in Eq.~\ref{eq:NJ}, we obtain that the total column density at the ortho state by:
\begin{equation}
    N_{T}=\frac{4 \pi F_J(T_{l}^{1-n}-T_{u}^{1-n})}{\Omega (\Delta E_{u-l}\,A_{u-l})(n-1)\int_{T_{u}}^{T_{l}} f(T) T^{-n}d T}.
\end{equation}
To estimate the gas mass, we need to determine the power-law index ($n$), the upper ($T_{u}$)  and lower ($T_{l}$) temperature limits. This can be done by dividing $N_{J}$ for the column density of a reference state, e.g. $N_3$, 
\begin{equation}
    \frac{N_{J}/g_{J}}{N_{3}/g_{3}}=\frac{ \int_{T_{u}}^{T_{l}}  e^{{-E_{J}}/{k_{B}T}} T^{-n}d T}{\int_{T_{u}}^{T_{l}} e^{{-E_{3}}/{k_{B}T}} T^{-n}d T}.
\end{equation}
Following \cite{2011ApJ...726...76Y}, we use  $T_u$\,=\,5000K to include the hot molecular phase, and vary the parameter space of $T_l$ and $n$ in order to minimize the equation above. The resulting models for the disk and outflow components are shown in Fig.~\ref{fig:temp}. The disk component is well reproduced with $n\,=\,4.84$ in a temperature range from 134 to 5000\,K, similarly to what was found by \citep{togi}, using Spitzer data. For the outflow, we find $n\,=\,4.26$ within a temperature range of 198$-$5000\,K. This shallower slope for the outflow component is consistent with shock heated gas cooling through the emission of the higher J transitions \citep{2017ApJ...836...76A}.

Finally, the mass of gas is obtained by
\begin{equation}
    M=(4/3) m_{\rm H_{2}} N_{T} \Omega D^{2},
\end{equation}
where $m_{\rm H_{2}}$ is the mass of the H$_2$ molecule and $\Omega D^{2}$ is the area of the encompassed region. The 4/3 factor is used to account for the mass of the molecule at the para state, considering a ortho to para ratio of 3. 
  We estimate the masses and outflow properties for the two phases the warm (198K-1000K) and the hot (1000\,K-5000\,K). We obtain masses of ($7.02\pm2.57) \times 10^{6} {\rm M_{\odot}}$ and ($1.95\pm0.71) \times 10^{5} {\rm M_{\odot}}$,  for the warm and the hot gas phases, respectively.

\begin{figure}
    \centering
    \includegraphics[width=.45\textwidth]{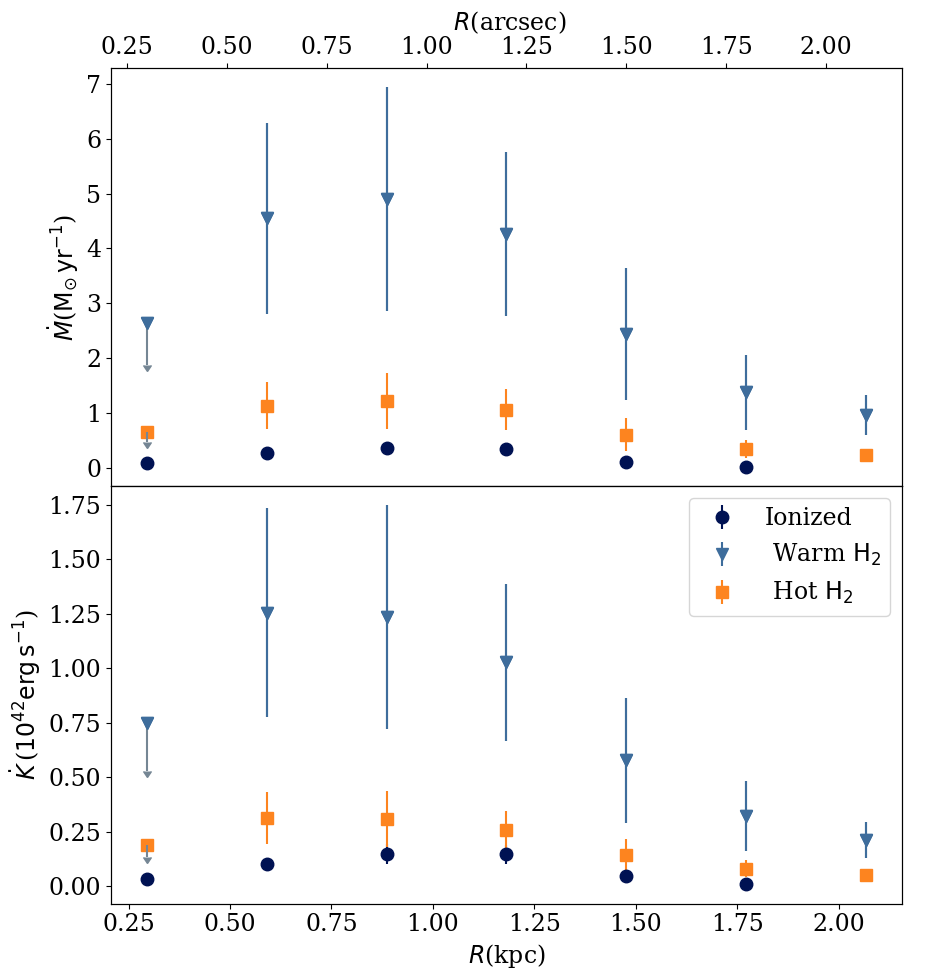}
    \caption{ Radial plots for the mass-outflow rate (top panel) and kinetic power of the outflows (bottom panel). The dark blue points represent the ionized outflows, warm H$_2$ is shown in light blue and the hot H$_2$ is shown as orange squares. The arrows in the first radial bin show the minimum values of the outflow properties, after correcting for the PSF scatter if an unresolved outflow is present. For bins further from the nucleus, the PSF contribution is negligible ($\lesssim$2\%), so the arrows are not shown.} 
    \label{fig:enter-fig4}
\end{figure}

To estimate the mass outflow rate ($\dot{M}$) and kinetic power ($\dot{K}$) of the outflow, we adopt a spherical shell outflow geometry, for which
\begin{equation}
    \dot{M} = \frac{M_{out,\Delta R}V_{out,\Delta R}}{\Delta R},
\end{equation}
where $M_{out,\Delta R}$ is the outflowing mass for each shell, with width of  $\Delta$R\,=\,0.3$\arcsec$ (close to the angular resolution), we also use the $V_{out,\Delta R}=v_{98} = |\langle V_{\rm wind}\rangle_{\Delta R}| + 2\langle \sigma_{\rm wind}\rangle_{\Delta R}$ as commonly done in the literature \citep{2013ApJ...768...75R,2021MNRAS.504.3890D}. The median velocity of the molecular outflows in UGC\,8782 is $\sim$900\:km\,s$^{-1}$. This value is smaller than the escape velocity of the galaxy, which is $\sim$1130\:km\,s$^{-1}$, implying that the outflows are only capable of redistributing gas within the galaxy. However, it will still be available for further star formation.

The kinetic power of the outflow for each shell is estimated as
\begin{equation}
    \dot{K} = \frac{M_{out,\Delta R}(V_{out,\Delta R})^{3}}{2\Delta R}.
\end{equation}

Fig.~\ref{fig:enter-fig4} presents the radial profiles of the mass-outflow rate (top panel) and kinetic power (bottom panel) for the warm  and hot molecular outflows, in bins of 0.3 arcsec. We also include, for comparison, the radial profiles for the ionized gas outflows based on the H$\alpha$ emission, using the data from \citet{2023MNRAS.521.3260R} and the same methodology described here, however with a seeing of $\sim$\,0.75 arcsec. The maximum outflow rate of $4.90\pm2.04$\,M$_{\odot}$yr$^{-1}$ is seen in the warm phase at $\sim$900\,pc from the nucleus, whilst the hot counterpart presents a lower  contribution, of $1.22\pm0.51$\,M$_{\odot}$yr$^{-1}$. {We estimate the contribution of scattered light from the nucleus to the derived mass outflow rates by scaling the flux of the broad component with the observed PSF for the star HD 2811, processed in the same way as the science data. We find a contribution of up to 30~\% for the central bin, and contributions lower than 2~\% for larger distances from the nucleus. The downward-pointing arrows in the first radial bin show the values after correcting for the PSF scatter if an unresolved outflow is present.}


Given the AGN bolometric luminosity of $L_{\rm bol}=(2.5-7.2)\times10^{43}$ erg\,s$^{-1}$ \citep{2023MNRAS.521.3260R}, the molecular outflow demonstrates maximum kinetic coupling efficiencies ($\epsilon$) ranging from 2--5\% of the AGN luminosity, the kinectic power of the warm and hot phase are respectively $1.26\times10^{42}$ erg\,s$^{-1}$ and $3.12\times10^{41}$ erg\,s$^{-1}$. This range far surpasses the minimum $\epsilon$ required for effective AGN feedback in suppressing star formation \citep{Hopkins10}, particularly when considering the potential presence of a cold molecular gas reservoir that may also contribute to the outflow.
 Nevertheless, the outflows in UGC\,8782 are likely mechanically driven by the radio jet, as the kinetic power of the outflow is one order of magnitude lower than the object's radio source power of $(2-4)\times$10$^{43}$ erg\,s$^{-1}$ \citep{lanz15}. Considering a constant outflow rate the warm and hot H$_2$ gas reservoir would be depleted in around $\sim 1$\,Myr.


 

\section{Conclusions} \label{sec:conclusion}

We reported the observations of warm molecular gas  spatially resolved outflows from the AGN of UGC\,8782. We find that the mass outflow rate in warm molecular gas is substantially higher than that observed in ionized gas across all radii, with coupling efficiencies exceeding the minimum required for AGN feedback to be effective in quenching star formation. The observed molecular outflows in UGC\,8782 of up to $\sim$6\:M$_{\odot}$yr$^{-1}$ are able of redistributing all the warm molecular gas available in the inner 2 kpc of the UGC\,8782 in only $\sim$1\,Myr.


\section*{acknowledgments}
We would like to express our gratitude to an anonymous referee for their valuable comments that helped us improve this work. Additionally, we extend our thanks to the e-Merlin support staff for their assistance in retrieving the data from their archive, and to Rob Beswick for providing additional data. HCPS and GLSO thank the financial support from Coordena\c c\~ao de Aperfei\c coamento de Pessoal de N\'ivel Superior - Brasil (CAPES) - Finance Code 001.  RAR acknowledges the support from Conselho Nacional de Desenvolvimento Cient\'ifico e Tecnol\'ogico (CNPq; Proj. 303450/2022-3, 403398/2023-1, \& 441722/2023-7), Funda\c c\~ao de Amparo \`a pesquisa do Estado do Rio Grande do Sul (FAPERGS; Proj. 21/2551-0002018-0), and CAPES (Proj. 88887.894973/2023-00). RR and TSB acknowledge the support from CNPq and FAPERGS. M.B. acknowledges funding support from NASA Astrophysics Data Analysis Program (ADAP) grant No. 80NSSC23K0750 and STScI grant JWST-GO-01717.001-A, which was provided by NASA through a grant from the Space Telescope Science Institute, which is operated by the Association of Universities for Research in Astronomy, Inc., under NASA contract NAS 5-03127. M.B. is also grateful to the IAU's Gruber Foundation Fellowship Program for the financial support.

\section*{Data Availability}
The data used in this work are part of the JWST cycle 1 project (ID 1928). The complete dataset can be accessed at the Mikulski Archive for Space Telescopes (MAST) platform at the Space Telescope Science Institute, through \dataset[10.17909/tazj-hp44]{https://doi.org/10.17909/tazj-hp44}.

\facilities{JWST (MIRI MRS), Gemini (GMOS-IFU) e-MERLIN, MAST, NED.}

\software{{\sc astropy} \citep{astropy:2022},
{\sc ifscube} \citep{Ruschel-Dutra21},
{\em JWST} Science Calibration \citep{bushouse_2024}, 
Matplotlib \citep{hunter07}, 
QFitsView \citep{ott12},
SciPy \citep{SciPy}.}


%

\appendix
\section{Beam Smearing}\label{beam}
The emission line profiles from distinct regions of a rotating disk component merge or blend together within the same resolution element, a effect known as beam smearing \citep[e.g.,][]{Swaters09}. Consequently, an unresolved disk component can produce a single observed emission line profile that appears to have contributions from multiple underlying velocity components. This effect occurs because the spatial resolution of the observation is insufficient to distinguish between the different line-of-sight velocities present across the disk. 

In order to evaluate the beam-smearing effect, we simulate a data cube in which the velocity field is given by a simple analytical model from \citet{1991ApJ...373..369B}, with a projected velocity amplitude of 200~km\:s$^{1}$ similar to the observed value. The $\rm H_{2}$\,S(5), 6.91$\mu$m emission line profile at each spaxel is represented by a Gaussian function with the flux decreasing with the distance from the nucleus and with a constant velocity dispersion of 100~km\:s$^{-1}$. The continuum emission is modeled by a power-law function, which decreases radially. Additionally, we include a random noise component that matches the observed noise level of the datacube.  Then, we convoluted the data cube with the JWST PSF, represented by the flux distribution of the star HD\,2811.

\begin{figure*}
    \centering
    \includegraphics[angle=0,scale=.45]{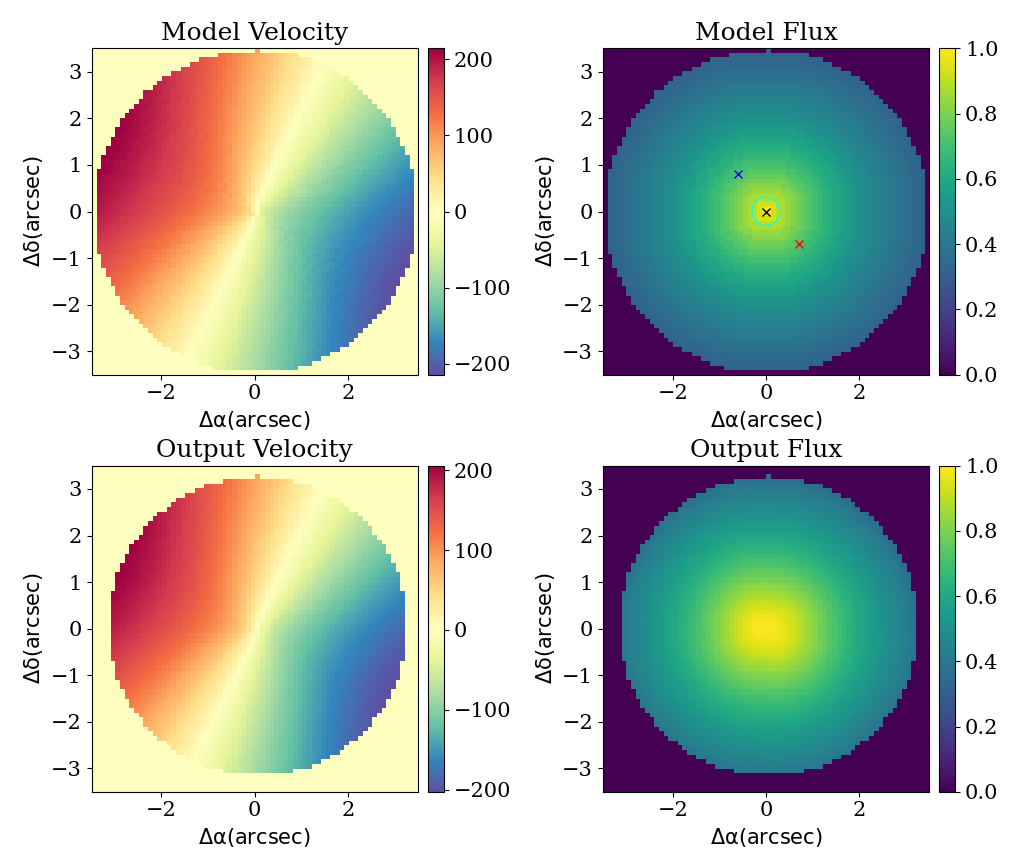}
    \includegraphics[angle=0,scale=.45]{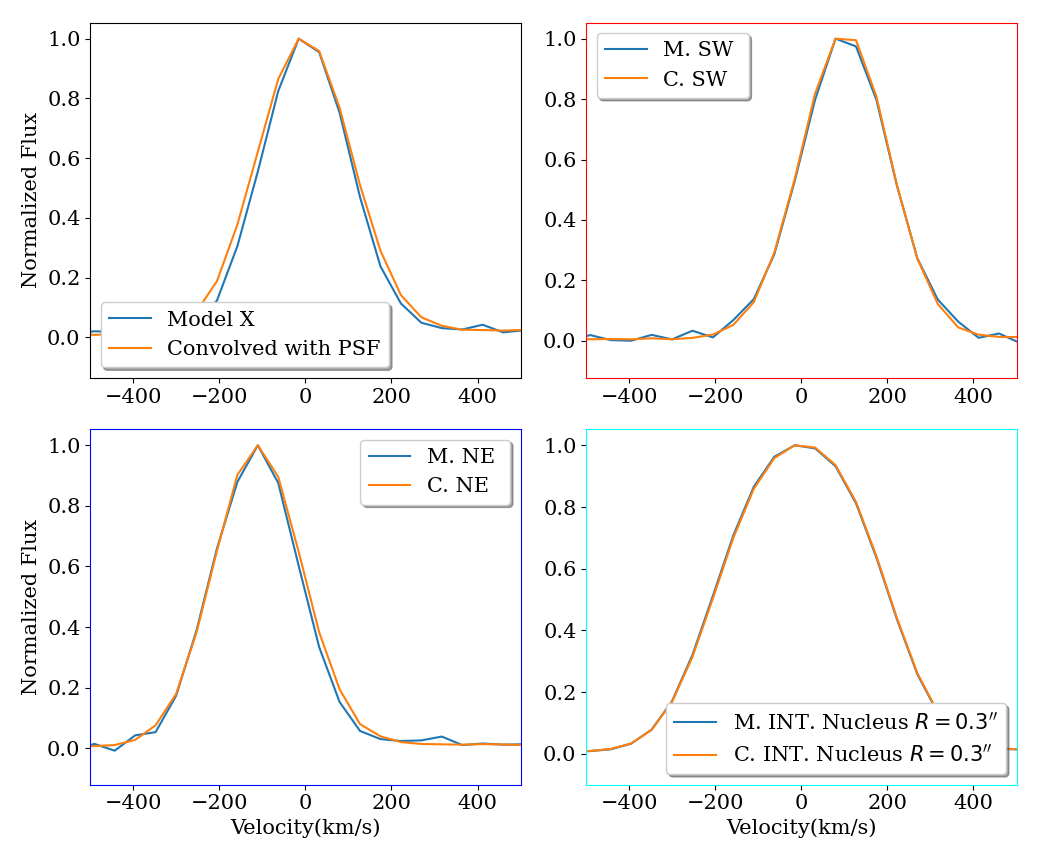}
    \caption{ The maps show the models for the velocity (left) and the emission-line flux distribution (right) before (top) and after (bottom) the convolution with PSF. Examples of line profiles before (in blue) and after (orange) the convolution with the PSF are shown at the bottom plots, for the positions indicated in the model flux map.}
    \label{fig:bms}
\end{figure*}

In Figure~\ref{fig:bms} we show the results of this toy model. The upper maps show the velocity field (left) and the emission-line flux distribution (right) before (top) and after (bottom) the convolution with the PSF, whilst the bottom plots show examples of the line profiles at the nuclear spaxel (top left), two spaxels along the major axis of the disk (top right) and (bottom left) and for an integrated aperture of 0.3 arcsec radius (bottom right). As can be observed from this figure, the line profiles before and after the convolution of the model with the PSF are similar, indicating that the effect of beam smearing is negligible.

\bibliography{sample631}{}
\bibliographystyle{aasjournal}



\end{document}